%% file: main.tex
\documentclass[conference]{IEEEtran}

\usepackage{cite}
\usepackage{amsmath,amssymb,amsfonts}
\usepackage{algorithmic}
\usepackage{graphicx}
\usepackage{textcomp}
\usepackage{xcolor}

\usepackage{url}
\usepackage{paralist}
\usepackage{subfiles}
\usepackage{listings}
\definecolor{codegreen}{rgb}{0,0.6,0}
\definecolor{codegray}{rgb}{0.5,0.5,0.5}
\definecolor{codepurple}{rgb}{0.58,0,0.82}
\definecolor{backcolour}{rgb}{0.95,0.95,0.92}
\lstdefinestyle{mystyle}{
    backgroundcolor=\color{backcolour},   
    commentstyle=\color{codegreen},
    keywordstyle=\color{magenta},
    numberstyle=\tiny\color{codegray},
    stringstyle=\color{codepurple},
    basicstyle=\ttfamily\footnotesize,
    breakatwhitespace=false,
    language=Java,
    breaklines=true,                 
    captionpos=b,                    
    keepspaces=true,                 
    numbers=left,                    
    numbersep=5pt,                  
    showspaces=false,                
    showstringspaces=false,
    showtabs=false,                  
    tabsize=2
}

\def\BibTeX{{\rm B\kern-.05em{\sc i\kern-.025em b}\kern-.08em
    T\kern-.1667em\lower.7ex\hbox{E}\kern-.125emX}}

\begin{document}

\title{Measuring Software Testability via Automatically Generated Test Cases}

\author{
\IEEEauthorblockN{1\textsuperscript{st} Luca Guglielmo}
\IEEEauthorblockA{\textit{Department of Informatics, Systems and Communication} \\
\textit{University of Milano-Bicocca}\\
Milano, Italy \\
luca.guglielmo@unimib.it}
\and
\IEEEauthorblockN{2\textsuperscript{nd} Leonardo Mariani}
\IEEEauthorblockA{\textit{Department of Informatics, Systems and Communication} \\
\textit{University of Milano-Bicocca}\\
Milano, Italy \\
leonardo.mariani@unimib.it}
\and
\IEEEauthorblockN{3\textsuperscript{rd} Giovanni Denaro}
\IEEEauthorblockA{\textit{Department of Informatics, Systems and Communication} \\
\textit{University of Milano-Bicocca}\\
Milano, Italy \\
giovanni.denaro@unimib.it}
}

\maketitle

\begin{abstract}
Estimating software testability can crucially assist software managers to optimize test budgets and software quality.
In this paper, we propose a new approach that radically differs from the traditional approach of pursuing testability measurements based on software metrics, e.g., the size of the code or the complexity of the designs. Our approach exploits automatic test generation and mutation analysis to quantify the evidence about the relative hardness of developing effective test cases. In the paper, we elaborate on the intuitions and the methodological choices that underlie our proposal for estimating testability, introduce a technique and a prototype that allows for concretely estimating testability accordingly, and discuss our findings out of a set of experiments in which we compare the performance of our estimations both against and in combination with traditional software metrics. The results show that our testability estimates capture a complementary dimension of testability that can be synergistically combined with approaches based on software metrics to improve the accuracy of predictions.
\end{abstract}


\section{Introduction}
\subfile{sections/introduction}
\section{Evidence-based Testability Estimation}\label{sec:approach}
\subfile{sections/technique} 
\section{Experiments}\label{sec:experiments}
\subfile{sections/experiments}
\section{Related work}\label{sec:related_works}
\subfile{sections/relatedWorks}
\section{Conclusions}\label{sec:conclusions}
\subfile{sections/conclusions}

\bibliographystyle{IEEEtran}
\bibliography{bibliography}

\end{document}

%% file: sections/introduction.tex
Software testing is a key activity of the software life-cycle that requires time and resources to be effective.
In this paper we focus on the \emph{testability} of the software, which is defined as \emph{the degree to which the design of software artifacts supports or hardens their own testing}~\cite{InternationalStandardOrganization(ISO).2001,IEEE1990}, and which can correlate in many relevant ways with the cost of the testing activities and ultimately with the effectiveness of those activities for revealing the possible faults.
For example, the availability of estimates on the testability of the software under test and the components therein can support test analysts in
anticipating the cost of testing, tuning the test plans, or pinpointing components that should undergo refactoring before testing.

At the state of the art the problem of estimating software testability has been addressed with two main classes of approaches: \emph{fault-sensitivity approaches}, which estimate testability by focusing on the probability of executing and revealing possible faults, and approaches based on \emph{software metrics}, which conjecture the correlation between the testing effort and the static structures of the code  characterized with software metrics as, for example, the cyclomatic complexity or the lines of code.

The \emph{fault-sensitivity approach} grounds on the seminal work of Voas and colleagues on the \textit{execute-infect-propagate} (PIE) model of fault sensitivity~\cite{voas_improving_1991,voas_predicting_1991,voas_pie_1992,voas_software_1995,bologna_object-oriented_1996}. The PIE model defines fault sensitivity as the combined probability of executing faulty locations, infecting the execution state and propagating the effects of the infection to some observable output. 
High fault sensitivity can be a proxy of high testability, and vice-versa.
However, doing actual estimates requires to observe the frequency of execution of the program locations with very thorough test suites, hardly available before testing~\cite{voas1993empirical}. As a matter of fact, after the initial momentum in the nineties, this approach has never made its way to established testability estimation tools and has been progressively abandoned.

The \emph{software metrics approach} is the main focus of most past and recent research on software testability~\cite{bruntink_predicting_2004,gupta_fuzzy_2005,bruntink_empirical_2006,singh_predicting_2008,badri_exploring_2010, khalid_analysis_2010,singh_predicting_2010,badri_empirical_2011,badri_empirical_2012,zhou_-depth_2012,da_cruz_empirical_2017,toure_predicting_2018,terragni_measuring_2020,alshahwan_improving_2009,khan_metric_2009,kout_empirical_2011}.

Most research effort focuses on object-oriented programs, by using metrics that capture information about the static structure of the code at the class-level or method-level (as for example the Chidamber and  Kemerer's metrics~\cite{chidamber_metrics_1994}). The software metrics that have the potential of being good testability predictors are derived by investigating the correlation between the metrics and the amount, the complexity and the thoroughness of the associated test cases.

We observe that a potential threat to the way these software metrics have been investigated is the fact that many of these studies are performed only on a, oftentimes small, sample of projects, and this could lead to generalization problems.
For instance, several studies report contrasting results: Bruntink et al. \cite{bruntink_predicting_2004, bruntink_empirical_2006} do not identify WMC and LCOM as good predictors differently to other studies \cite{badri_empirical_2011, badri_empirical_2012, gupta_fuzzy_2005, singh_predicting_2008, da_cruz_empirical_2017} and NOC is identified as a good predictor only by Singh et al. \cite{singh_predicting_2008}, while others have not found such correlation \cite{bruntink_predicting_2004, bruntink_empirical_2006,badri_empirical_2012,terragni_measuring_2020}.

In this paper we introduce and discuss the novel idea of not relying on the possible correlation between static metrics and testability, but to \emph{directly estimate the testability degree of a software by sampling the test space and the fault space} of the software, and therein collect empirical \emph{evidence of the easiness or hardness} to accomplish effective testing.
According to our approach, the stronger the evidence that we can collect about hard-to-test faults in a software component, the higher the probability that its design is not facilitating testing enough.
Drawing on this idea, we rely on a search-based test generation tool to automatically generate test cases~\cite{fraser_evosuite_2011}, and refer to mutation-based fault seeding to sample possible faults~\cite{coles_pit_2016}. We then refer to the generated test cases and the seeded faults to  \emph{extrapolate the testability evidence}.

We empirically studied the effectiveness of our testability estimates with respect to 598 class methods of three large software projects in Java. In particular we analyzed to what extent our estimates correlate with the development complexity of the test cases that were available in the considered projects, and we compared the correlation yielded by our estimates with the one yielded by a selection of popular software metrics for object-oriented programs. 
Our main findings were that our testability estimates contribute to explain the variability in the development complexity of the test cases by capturing a different phenomenon than the metrics on the size and the structure of the software.
Furthermore, motivated by such findings, we explored the combination of our metric with the software metrics, revealing synergies to improve the testability estimates.
Thus, our findings support the research hypothesis that it is viable and useful to estimate testability based on empirical observations collected with automatically generated test cases. We remark that we do not claim that  testability estimates based on software metrics must be avoided and replaced with our testability estimates, but rather the two approaches could be used synergically to improve the accuracy of the estimates.

The paper is organized as follows. 
Section \ref{sec:approach} presents our novel approach to estimate testability. Section \ref{sec:experiments} presents our experiments.
Section \ref{sec:related_works} surveys the relevant related work. Section \ref{sec:conclusions} summarizes the main contributions of this paper.

%% file: sections/technique.tex
 Providing testability measurements amounts to estimating \emph{the degree to which a software component facilitates its own testing}~\cite{freedman_testability_1991,voas_pie_1992,IEEE1990,InternationalStandardOrganization(ISO).2001}.
In this section we elaborate on both the intuitions and the  methodological choices that underlie our novel proposal to make these estimates.

\subsection{Intuitions and  Approach Overview}

\subsubsection{Exploiting automated test generation}
The main intuition that inspires our approach is to experience with the testability of a given piece of software by simulating the activity of crafting test cases for that software. Namely, we rely on automatically generating test cases with a test generator (in this paper we used the test generator Evosuite~\cite{fraser_evosuite_2011}).
Looking into the results from the test generator, we aim to judge the extent to which the current design is making it hard (or easy) for the test generator to accomplish test objectives against the considered software.

\subsubsection{Exploiting mutation analysis} 
We sample  possible test objectives in the form of synthetic faults injected in the target software.
We rely on \emph{mutation-based fault seeding}~\cite{demillo_hints_1978,pezze_software_2008}. 

Mutation-based fault seeding injects possible faults by referring to so-called mutation operators, each describing a class of code-level modifications that may simulate faults in the program.
For instance, the mutation operator \emph{replace-arithmetic-operators} creates faulty versions (called mutants) of a program by exchanging an arithmetic operator used in the code with a compatible arithmetic operator: it can produce a faulty version for each possible legal replacement.
In the sample Java program in Figure~\ref{fig:sample_program} we indicated some possible mutants in the comments included in the code: For example, we can create a faulty version of the program by replacing the statement at line~\ref{fig:sample_program:mut_easy} with the statement indicated in the comment at the same line. This is a possible instance of \emph{replace-arithmetic-operators}; another one is the mutant indicated in the comment at line~\ref{fig:sample_program:mut_not_observable}. The mutants indicated at lines~\ref{fig:sample_program:mut_not_baseline} and~\ref{fig:sample_program:mut_hard} refer to another possible mutation operator, \emph{replace-expressions-with-literals}, which consists in replacing a numeric expression with a compatible constant mentioned somewhere else in the code.
In this paper we used the mutation analysis tool PIT~\cite{coles_pit_2016}.

\begin{figure}[t!]
\begin{lstlisting}[language=Java,escapechar=@]
public class SampleProg {
	public static finale int TARGET = 10;
	private int state;
	private int scale;
	private int mode;
	private int sensor;
	public void setScale (int s) {
		scale = s * 100; //MUT: scale = s + 100 @\label{fig:sample_program:mut_easy}@
	}
	public void setMode(int m) {
		if (new fileExists("/config.conf")) {
			int conf = read("/config.conf"); 
			mode = conf; //MUT: mode = 100 @\label{fig:sample_program:mut_not_baseline}@
		} else mode = m;
	}
	public void setSensor(int s) {
		sensor = s;
	}
	public int getScale() {
		GUI.msg(scale / 10); //MUT: GUI.msg(scale*10) @\label{fig:sample_program:mut_not_observable}@
		return scale;
	}
	public void updtState(){
		if (mode == TARGET) peekSensor();
		else state = 1;
	}
	private void peekSensor(){
		if (scale > 1000000) state = sensor;
		else state = abs(sensor);
	}
	public int currState() {
		if (state >= 0) return state;
		else return -1; //MUT: return 1 @\label{fig:sample_program:mut_hard}@
	}
}
\end{lstlisting}
\caption{A sample Java programs and some corresponding mutants}
\label{fig:sample_program}
\end{figure}

To judge testability, we focus on each seeded fault separately, and we evaluate whether the current program, by virtue of its design, makes it hard (or easy) for the test generator to reveal that fault: If the test generator succeeds to reveal the fault, 
we infer a piece of testability evidence, under the intuition that a human could succeed as well with controlled effort; Otherwise, we might infer a non-testability evidence (though this  requires the further analysis described below,~$\S$~\ref{sec:custom_setters}).

For instance, our prototype based on Evosuite easily reveals the mutant at line~\ref{fig:sample_program:mut_easy} of  Figure~\ref{fig:sample_program}, e.g., with a test case like
\begin{lstlisting}[language=Java,numbers=none]
SampleProg p = new SampleProg();
p.setScale(0);
assertEquals(0, p.getScale();
\end{lstlisting}
whose execution fails against the mutant, but not for the original program. We thus infer 
a piece of testability evidence after observing that Evosuite easily reveals this mutant. 
Conversely, revealing the mutant at line~\ref{fig:sample_program:mut_hard} requires  a test case carefully tuned on several interdependent class methods, which Evosuite consistently fails to generate. A test case that could reveal this mutant would be resemblant to the following one,
\begin{lstlisting}[language=Java,numbers=none]
SampleProg p = new SampleProg();
p.setMode(10); // Hit mutant iff this.state = 10...
p.setScale(20000); // ...and this.scale > 1000000...
p.setSensor(-5); // ...and this.sensor < 0 
p.updtState(); // ...when executing updtState().
assertEquals(-1, p.currState();
\end{lstlisting}
which requires an arguably non-negligible amount of mental and manual effort also for a human tester. We infer a piece of non-testability evidence after observing that the test generator is unable to reveal this latter mutant.

Eventually, we aggregate the testability and non-testability evidence across the seeded faults of the piece of software of interest, to estimate the degree of testability of that software: The larger the amount of testability (resp. non-testability) evidence, i.e., many mutants are easy (resp. hard) for the test generator to reveal, the higher (resp. lower) the estimated degree of testability.

\subsubsection{Exploiting testability-facilitated APIs as baseline}
\label{sec:custom_setters}

When inferring non-testability evidences as above, we must pay attention that the quality of our estimations could be jeopardized by intrinsic limitations of the approaches (and ultimately the tools) to which we refer for generating the test cases.
In particular, we aim to \emph{avoid non-testability judgements that can derive from intrinsic limitations of the  test generator}, rather than testability issues.

For example, a test generator that is not able to construct some types of data structures, or does not handle test data from files or network streams, will systematically miss test cases for any fault that depends on those types of test data, regardless of actual testability issues of the software under test.
Evosuite consistently fails to hit the mutant at line~\ref{fig:sample_program:mut_not_baseline} of Figure~\ref{fig:sample_program}, simply because manipulating the file system (to set a proper file \texttt{/config.conf}) is not part of its functionality.

To acknowledge cases of this type, our approach constructively discriminates the subset of seeded faults (out of the ones provided by the mutation-analysis tool) for which we can acquire a \emph{sufficient evidence} (not necessarily a proof) that they are not out of the scope of the considered test generator.
We refer to the resulting subset of mutants as the \emph{baseline mutants}, since they provide the actual baseline for us to judge the testability evidences.

In our approach, the baseline mutants are those that the test generator either can already reveal or could reveal if it has the freedom to both directly assign any state variable in the program, and directly inspect the infected execution states. 
We enable this capability by 
\begin{inparaenum}[(i)]
    \item augmenting the program under test with custom setters for all the state variables of any module in the program, and
    \item recording the mutants as revealed-mutants as soon as they get executed, even if there is not a failing assertion in the test cases.
\end{inparaenum}
We refer to this simplified setting of the testing problem that we submit to the test generator as \emph{testing the program with testability-facilitated APIs}.

For example, in the case of the program in Figure~\ref{fig:sample_program}, we equip the program with the following set of custom setters:
\begin{lstlisting}[language=Java,numbers=none]
public void _custom_1__(int i) {this.state = i;}
public void _custom_2__(int i) {this.scale = i;}
public void _custom_3__(int i) {this.mode = i;}
public void _custom_4__(int i) {this.sensor = i;}
\end{lstlisting}

In the simplified setting, Evosuite easily executes 
\begin{inparaenum}[(i)]
    \item the mutant at line~\ref{fig:sample_program:mut_easy}, e.g., with the test case that we already discussed above, 
    \item the mutant at line~\ref{fig:sample_program:mut_hard}, e.g., with a test case like
\begin{lstlisting}[language=Java,numbers=none]
SampleProg p = new SampleProg();
p._custom_1__(-3); // Sets state with custom setter
assertEquals(-1, p.currState();
\end{lstlisting}
that suitably exploits the custom setter \texttt{custom\_1\_\_} to workaround the \emph{hard-for-testing API} of the original program for controlling a negative value of \texttt{this.state},
and \item  the mutant at line~\ref{fig:sample_program:mut_not_observable}, e.g., with a test case like
\begin{lstlisting}[language=Java,numbers=none]
SampleProg p = new SampleProg();
p.getScale();
\end{lstlisting}
that makes the mutant generate an infected execution state at line~\ref{fig:sample_program:mut_not_observable}, even if Evosuite cannot generate a proper assertion for it.
\end{inparaenum}
However Evosuite cannot anyway hit the mutant at line~\ref{fig:sample_program:mut_not_baseline} that remains out of the scope of the test generator, regardless of the availability of the custom setters.
Thus, we eventually consider as baseline mutants only the mutants at lines~\ref{fig:sample_program:mut_easy}, \ref{fig:sample_program:mut_not_observable} and~\ref{fig:sample_program:mut_hard}, but not the one at line~\ref{fig:sample_program:mut_not_baseline}. 

In summary, our testability judgements are made by generating test cases for both the original program and the program augmented with the custom setters, and mutually crosschecking both sets of test results. We infer testability evidences upon observing that the test generator successfully generates test cases that reveal mutants in the original program, e.g., the mutant at Figure~\ref{fig:sample_program}, line~\ref{fig:sample_program:mut_easy}. We infer non-testability evidences out of the inability of the test generator to reveal baseline mutants, e.g., the mutants at lines~\ref{fig:sample_program:mut_hard} and~\ref{fig:sample_program:mut_not_observable}. But our estimations dismiss the information about the non-baseline mutants (as the one at line~\ref{fig:sample_program:mut_not_baseline}) conjecturing that the test generator could not address those mutants regardless of the testability of the program.

We are aware that, technically speaking, using the testability-facilitated APIs may lead us to generate some input states that are illegal for the original program. Nonetheless, we embrace this approach heuristically: observing faults that the test generator can hit only with the testability-facilitated APIs suggests restrictive designs of the program APIs, which may pinpoint testability issues.

\subsection{The Technique}\label{sec:technique}
\begin{figure*}[t!]
\centerline{\includegraphics[width=0.90\textwidth]{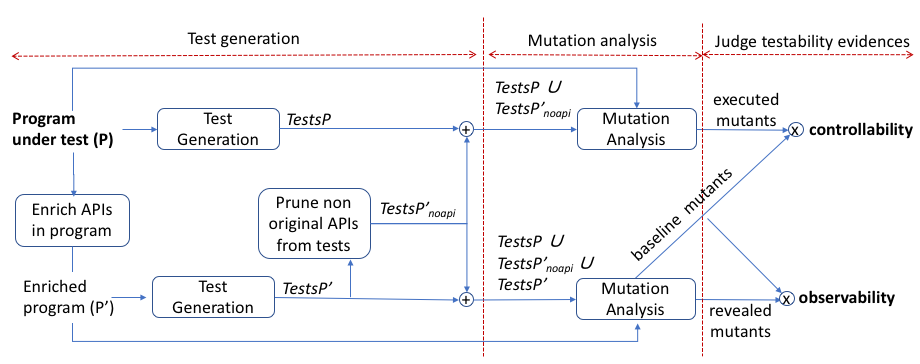}}
\caption{Workflow of our technique to estimate testability evidences}
\label{fig:technique}
\end{figure*}

Figure~\ref{fig:technique} illustrates the workflow by which our technique exploits automated test generation (left part of the figure) and mutation analysis (middle part of the figure) in order to judge testability evidences (right part of the figure). 

The input is a given program under test, which is indicated at the top-left corner in the figure, and the result is a set of testability evidences, classified as either controllability evidences or observability evidences, as indicated at the rightmost side of the figure. The blocks named \textit{Test Generation} indicate test generation activities.
The blocks named \textit{Mutation Analysis} indicate mutation analysis activities.
The block named \textit{Enrich APIs in program} augments the program under test with the testability-facilitated APIs, as we introduced in the previous section. The block \emph{Prune non original API from tests} removes the calls to the testability-facilitated APIs from the test cases generated for the program augmented with the testability-facilitated APIs, to obtain additional test cases (and thus further testability evidence) for the original version of the program, as explained below.
The circles that contain the \textit{+} symbol indicate post-processing for merging the generated test suites into a single test suite. The circles that contain the \textit{$\times$} symbol indicate post-processing of the data derived from mutation analysis to derive controllability and observability evidences.
The arrows specify the inputs and the outputs of each activity.

We have currently implemented the entire workflow of Figure~\ref{fig:technique} for programs in Java as a fully automated process scripted in Bash and Java.
Below we explain all details of our approach for the three phases of the workflow.

\subsubsection{Test Generation}
Our current implementation generates test cases with the test generator EvoSuite that exploits a \emph{search-based test generation algorithm} to generate test cases for Java classes~\cite{fraser_evosuite_2011}. 
Given a Java class and a set of code coverage criteria, EvoSuite starts with randomly sampling a first set of possible test cases for the class, and then iterates through evolving the test cases multiple times by applying random changes, while searching for sets of test cases that optimize the given code coverage criteria.
Furthermore, it generates assertion-style oracles on the observed outputs.
 
With reference to the \textit{Test Generation} blocks in Figure~\ref{fig:technique}, our technique runs EvoSuite against both the program under test and its augmented version \textit{P'}. 
Our implementation of the block \textit{Enrich APIs in program} obtains the augmented program \textit{P'} by enriching the interfaces of all classes with custom setters for any class variable declared in the code.
In Figure~\ref{fig:technique} we denoted as \textit{TestsP} and \textit{TestsP'} the test suites generated as result of those EvoSuite runs, respectively.
The test suite \textit{TestsP'} generated against \textit{P'} indicates  program behaviors that EvoSuite could provably exercise, possibly with the help of facilitated APIs. At the same time, the test suite \textit{TestsP'} implicitly captures the program behaviors that the test generation algorithm of EvoSuite is unlikely to exercise, since it failed even when facilitated by the capability to set the input state independently from the constraints encoded in the program APIs.
The comparison between the test suites \textit{TestsP'} and \textit{TestsP} indicates program behaviors that arguably were hard to exercise due specifically to the constraints encoded in the APIs, that is, behaviors that do not belong to  \textit{TestsP} while being in \textit{TestsP'}.

For each of the \textit{Test Generation} blocks in Figure~\ref{fig:technique}, our technique runs EvoSuite for a maximum time budget that depends on the size of the class, considering a minimum time budget of two minutes for the smallest classes in the considered project and a maximum time budget of 20 minutes for the largest classes, while linearly scaling the time budget for the classes of intermediate size.

Furthermore, our technique acknowledges the dependency of the search-based algorithm of EvoSuite from the different code coverage criteria that the tool allows as possible fitness functions, and from the intrinsic randomness that can naturally make EvoSuite generate different sets of test cases at different runs. Aiming to exercise as many program behaviors as possible, we set EvoSuite to address  all available fitness functions, i.e, line coverage, branch coverage, output coverage, exception coverage, and mutation coverage.

To get rid of the confounding effect of the differences between the test suites \textit{TestsP} and \textit{TestsP'} that might be just due to randomness, we constructively merge those test suites as follows.
We test the original program with both the test cases from \textit{TestsP} and the ones from \textit{TestsP'$_{noapi}$}, i.e., the test cases that either were generated in \textit{TestsP'} but still did not use any custom setter, or could be adapted from test cases in \textit{TestsP'} by commenting the calls to the custom setters (Figure~\ref{fig:technique}, block \textit{Prune non original APIs from tests})\footnote{\textit{TestsP'$_{noapi}$} represents test cases that Evosuite could generate also for the original program. Note that the custom setters can be safely commented without breaking the syntactic validity of the test cases.}. 
For similar reasons, all the available test cases must be accounted among the ones that EvoSuite could generate for \textit{P'}, that is, \textit{TestsP}, \textit{TestsP'} and \textit{TestsP'$_{noapi}$}.

\subsubsection{Mutation Analysis}
We use the mutation analysis tool PIT to both seed possible faults of the program under test~\cite{coles_pit_2016}, and characterize the generated test suites according to their ability to execute and reveal those seeded faults. 
This is the information that we will use in the next phase to judge the testability evidences that the generated test suites provided for the program under test.

PIT seeds faults in the program under test according to the mutation operators described in the documentation of the tool.\footnote{\url{https://pitest.org/quickstart/mutators/}} We specifically considered the set of mutation operators that PIT advises as the "stronger" group, which includes 13 mutation operators that address several types of  mutations at the level of the arithmetic operators, the comparison operators, the logic operators, the return values and the if and switch statements in the programs\footnote{We did not consider the larger set of mutation operators that PIT refers to as the "all" group because either they are marked as \emph{experimental} in the documentation, or our initial experiments showed that they result most often in duplicating mutants that we already obtain with the operators of the "stronger" group.}. 

PIT monitors the execution of the test suites against the mutants that it computes according to the selected mutation operators, and classifies the mutants as either \emph{revealed}, \emph{executed} or \emph{missed}. PIT classifies a mutant as revealed, if at least a test case produces a different result when executed against the original program or the mutant program, respectively. That is, (i) the test case executes with no exception and raises an exception for either program, or (ii) it raises different exceptions for either program, or (iii) it passes all test oracles and fails for at least a test oracle for either program, or (iv) it fails with respect to different test oracles  for either program.
Our technique considers the  assertion-style oracles that EvoSuite generated in the test cases.\footnote{We instructed EvoSuite to generate assertions for all program outputs encompassed in the test cases (that is with the option \texttt{assertion\_strategy=ALL}), since we aim to reveal as many mutants as possible, even if the test cases could become large.} PIT classifies a mutant as executed, if it could not classify the mutant as revealed, but there is at least a test case that executes the code in which the corresponding fault was injected.
PIT classifies a mutant as missed if it could not classify it neither as revealed nor as executed. 

In Figure~\ref{fig:technique}, the two blocks \textit{MA} indicate that our technique executes PIT for the test suites that we generated with EvoSuite for both the program under test and its augmented version\footnote{When running PIT on \textit{P'}, we do not inject mutations in the API methods that we artificially added to obtain exactly the same set of mutants for both programs \textit{P} and \textit{P'}.}. As result we collect:
\begin{itemize}
    \item the \textit{baseline mutants}, i.e., the mutants that are executed with the test cases run against for \textit{P'}. 
    \item the \textit{executed mutants}, i.e., the mutants that are executed with the test suite generated for the original program. These mutants were provably executed with EvoSuite with the original program APIs.
    \item the \textit{revealed mutants}, i.e., the mutants that are revealed with the test suite generated for \textit{P'}. These mutants were provably revealed with actual assertions within at least a test case in which they could be successfully executed.
\end{itemize}

\subsubsection{Testability Evidences}
Based on the results of mutation analysis, we look for indications of the testability of the program under test. Specifically, we first judge the testability of the program with respect to the testing goal that each mutant represents: do the results provide evidence that \emph{the program under test facilitates its own testing} with respect to the goal of revealing the seeded fault that each mutant represents?
By answering yes or no to this question we infer a testability evidence or an  evidence of non-testability, respectively, for each specific mutant that belongs to the set of baseline mutants computed as above.

We further split the testability verdicts into controllability verdicts and observability verdicts. Controllability refers to whether or not the results of mutation analysis provide evidence that the program under test facilitates the execution of the seeded faults. We annotate a controllability evidence for each mutant that mutation analysis marks as \textit{executed} for the original program under test, i.e., the set of executed mutants computed as above. 
With respect to these mutants, the test cases that we generated with EvoSuite provide empirical evidence that the program under test, with its original APIs, provides sufficient means of controlling the assignment of the program inputs and the program states for test cases to achieve the execution of those seeded faults. 
On the other hand, we annotate a non-controllability evidence for each baseline mutant not marked as executed for the original program. 

Observability refers to whether or not the results of mutation analysis provide evidence that the program under test facilitates to reveal the seeded faults. We annotate an observability or non-observability evidence for each baseline mutant that mutation analysis marks or does not mark, respectively, as \textit{revealed}.
The observability evidences correspond to empirical evidence that the program under test provides sufficient means for the seeded faults to be observed from the test cases.  

We aggregate the testability evidences, i.e., both the controllability and the observability evidences, for the mutants that correspond to faults seeded at the same line of code, to prevent the unbalanced skewing of our results towards those instructions that were associated with higher numbers of mutants than other instructions.
For each line of code associated with at least a baseline mutant, we infer a unitary controllability (resp. observability) evidence if more than half of the associated baseline mutants vote as controllability (resp. observability) evidences; or we infer unitary non-controllability (resp. non-observability) evidence otherwise. 

\subsection{Estimating Testability}
We refer to the collected testability and non-testability evidences to reason on the testability of given parts (e.g., software components) of the program under test.
For instance, in the experiments of this paper, we aimed to estimate testability values that represent the  testability of the methods that belong to a Java program. 

To this end, we first map each target piece of software (e.g., each method) to the subset of testability evidences that relate with that software, and then aggregate those testability evidences into a testability value measured in the interval $[0, 1]$, where 0 and 1 correspond the minimum and the maximum testability values that we can estimate for a component, respectively. 
 
Let $C$ be a software component that belongs to the program under test, and let $contr^+(C)$, $contr^-(C)$, $obs^+(C)$ and $obs^-(C)$ be the subsets of positive and negative controllability and observability evidences, respectively, that we mapped to the component $C$, out of the unitary evidences collected with the technique that we described in the previous section. Then, by referring to the size of those sets, we estimate the controllability and the  observability of the component $C$ as:

\[Controllability(C) = \frac{|contr^+(C)|}{|contr^+(C)| + |contr^-(C)|},\]

\[Observability(C) = \frac{|obs^+(C)|}{|obs^+(C)| + |obs^-(C)|}.\]

\noindent Finally we estimate the  testability of the component $C$ as the combination of its controllability and its observability, namely, as the arithmetic product of the two:

\[Testability(C) = Controllability(C) \times Observability(C).\]

Furthermore, we acknowledge that the testability evidences collected with our technique can be sometimes insufficient to calculate reliable estimates for some program components. In particular, we reckon this to be the case if our technique was unable to significantly sample the execution space of the component.
When reasoning on the testability of a piece of software, we mark our estimates as \emph{inconclusive} if the portion of lines of code for which we successfully computed testability evidences was not a representative sample out of the component's lines of code that were associated with some mutants.
We ground on the classic theory of small sample techniques~\cite{krejcie_determining_1970}. 
As a consequence, the possibility of producing inconclusive results for some components is a possible limitation of our technique. Depending on the actual implementations of the technique, the concrete manifestation of this limitation boils down to the characteristics of the tools with which we instantiate the test generation tool and mutation analysis phases. Explicitly pinpointing the conclusiveness of the estimates aims to alleviate the impact of such limitation.

%% file: sections/experiments.tex
We investigated to which extent our estimates of software testability for the methods that belong to a Java program can capture the actual complexity of developing test cases for those methods, in a set of experiments with many class methods and test cases out of three large Java projects.

We remark upfront the foundational nature of our current experiments.  
In particular we do not make any strong claim on the efficiency of either our current implementation, or the test generator and the mutation analyzer that the current implementation depends on. 
The main goal of the experiments reported in this section is to explore if there is merit in our idea of estimating testability by relying on empirical observations made with automatically generated test cases, and the possible complementarity between this new approach and the traditional approach of relying on the correlation with static software metrics.

\subsection{Research questions}
Our experiments were driven by the following research questions:

\begin{inparaenum}[RQ1:]
\item How large is the portion of inconclusive estimates with our current implementation of the technique? 

\item To what extent do our (conclusive) testability estimates correlate with the development complexity of the test cases that were designed for the considered methods, and how do they compare with traditional software metrics in this respect? 

\item Does combining our testability estimates with other static metrics improve over using only the static metrics as predictors of testability?
\end{inparaenum}

\subsection{Subjects}
We selected from GitHub three open-source Java projects that
\begin{inparaenum}[(i)]
\item use Maven as build tool, as this is a requirement of our current implementation of the technique, 
\item are representative of large projects comprised of at least 500 classes,
\item include at least 300 methods that can be associated (with the procedure that we describe in Section~\ref{testMethodAssociation}) with  test cases available in the projects,  
\item are representative of different types of software developments, namely, a programming library, a software engineering tool and a business oriented application. 
\end{inparaenum}
The three projects are: 
\begin{itemize}
\item JFreeChart, a \emph{programming library} that supports the display of charts,
\item Closure Compiler, a \emph{software engineering tool} that parses and optimizes programs in Javascript, and
\item OpenMRS-Core, a \emph{business-oriented application} for the healthcare domain.
\end{itemize}

\begin{table*}[htbp]
\small
\centering
\caption{\label{tab:overallStats}Descriptive statistics of the subject methods in the three considered Java projects}
\begin{tabular}{r|ccc|ccc|ccc|}
 & \multicolumn{3}{c|}{\textbf{JFreeChart}} & \multicolumn{3}{c|}{\textbf{Closure Compiler}} & \multicolumn{3}{c|}{\textbf{OpenMRS}} \\
& All & Tested & Subjects & All & Tested & Subjects & All & Tested & Subjects \\
\hline
\textbf{Number of methods}& 8552 & 613 & 246 & 14723 & 300 & 111 & 9166 & 493 & 241 \\
\textbf{Total lines of code (LOC)}& 71703 & 6751 & 5162 & 110553 & 3195 & 2266 & 51412 & 6527 & 5402 \\
\textbf{Average LOC per method}& 8.38 & 11.01 & 20.98 & 7.51 & 10.65 & 20.41 & 5.61 & 13.21 & 22.41 \\
\textbf{Mininum LOC per method}& 1 & 3 & 5 & 1 & 1 & 1 & 1 & 3 & 6 \\
\textbf{Maximum LOC per method}& 288 & 188 & 188 & 433 & 246 & 246 & 221 & 121 & 121 \\
\textbf{Average mutants per method}& 2.45 & 3.69 & 7.42 & 2.85 & 3.64 & 7.59 & 1.33 & 4.70 & 8.28 \\
\textbf{Mininum mutants per method}& 0 & 1 & 3 & 0 & 1 & 3 & 0 & 1 & 3 \\
\textbf{Maximum mutants per method}& 104 & 74 & 74 & 617 & 66 & 66 & 74 & 46 & 46 \\
\end{tabular}
\end{table*}

Table~\ref{tab:overallStats} summarizes descriptive statistics about the Java methods that belong to each project, namely, the number of methods (first row), their total and individual size (from the second to the fifth row), and the number of mutants in the methods (from the sixth to the eighth row).
The columns \textit{All} refer to all methods in the projects, while the columns \textit{Tested} and \textit{Subjects} refer, respectively, to the subset of methods that we were able to successfully associate with some test cases, and to the further subset that we selected as actual subjects for our study. We describe the procedure by which we selected these two latter subsets in the next section.
The data in the table indicate that we selected methods with increasing size and increasing number of mutants at each selection step.

\subsection{Ground Truth} \label{testMethodAssociation}
Out of the Java methods in the considered projects (Table~\ref{tab:overallStats}, columns \textit{All}), we excluded all methods \textit{hashCode} and \textit{equals} that are usually generated automatically, and further selected only the methods that we could associate with a reference ground-truth, that is, available test cases that the programmers developed for those methods.
This because we aimed at investigating the correlation between our testability estimates for the methods and the development complexity of the corresponding test cases, for methods and test cases designed by human programmers.
We built on the \textit{methods2test} tool~\cite{tufano_methods2test_2022} to associate the methods with the test cases available in the projects, and selected only the methods for which we identified at least an associated test case (Table~\ref{tab:overallStats}, columns \textit{Tested}).

Methods2test heuristically infers the associations between the available test cases and the methods that are their main testing target. It originally relies on two heuristics, \emph{name matching} and \emph{unique method call}, but we extended it with three additional heuristics, \emph{stemming-based name matching}, \emph{contains-based name matching} and \emph{non-helper unique method call},
which generalize the two original ones with the aim to increase the set of identified associations. 

For each test case, which in the considered projects is a test method within a test class, \emph{name matching} searches for a target method that both exactly matches with the name of the test case  and belongs to a class that exactly matches with the name as the test class.
\emph{Stemming-based name matching} and \emph{contains-based name matching} address the name matching with respect to either the stemmed names of methods and test cases, or  whether the test name contains the method name, respectively. For example \textit{testCloning}  and \textit{testCloneSecondCase} will match with method \textit{clone} after name stemming or by name containment, respectively.  
\emph{Unique method call} further exploits the name-based association between a test class and a target class, by searching for test cases that call a single method of the target class. \emph{Non-helper unique method call} re-evaluates the unique-method check after excluding the calls to possible helper methods, such as setter methods, getter methods and the method \textit{equals}.

After the association with the test cases, we further refined the set of subject methods by excluding the methods for which PIT computed mutants for at most two lines of code. For these methods our technique could  distill the unitary testability evidences out of a too squeezed population of seeded faults, which results in yielding unbalanced estimates in most cases. 
We see this as a drawback of the fault models that we are currently able to consider by relying on PIT, rather than as a limitation of our idea of estimating testability based on automatically generated test cases, and we thus dismissed these methods from the current experiments on this basis. We ended with selecting the set of subject methods summarized in the columns \textit{Subjects} of Table~\ref{tab:overallStats}.

We quantified the development complexity of the test cases associated with each subject method as the number of unique method invocations made within the ensemble of those test cases (counted with the tool CK) . We refer to this values as \textit{RfcTest}, i.e., the \textit{Rfc} values of the test cases.
Since the test cases are often sheer sequences of method calls (no decisions, no loops) other complexity metrics (like the cyclomatic complexity) are scarcely representative, while size metrics (like LOC, number of test cases or number of assertions) are more sensitive than RfcTest to arbitrary choices of testers.
\textit{RfcTest} represents more consistently than other metrics the effort that testers spent for understanding methods of other classes, as also considered in several testability studies \cite{badri_exploring_2010, badri_empirical_2011, kout_empirical_2011}.

Furthermore, we assessed the reliability of our  ground-truth with respect to possible errors in the method-test associations returned by Methods2test, by manually crosschecking 10\% of the methods (randomly sampled with R's function \textit{sample}) for which our technique produced a conclusive estimation (cfr.\ Section~\ref{sec:RQ1}, Table~\ref{tab:nomCovered}). 
Out of 42 subject methods that we crosschecked, we detected need for corrections for 7 methods, i.e., 5 methods for which Methods2test reported a wrongly matched test case (false positives), and 2 other methods for which Methods2test missed 3 and 2 associations, respectively (false negatives). For 3 of these 7 methods, correcting the errors of Methods2test did not affect the \textit{RfcTest} value, thus the corrections impacted only 4 out 42 methods. This datum suggests mild impact of the possible errors of Methods2test.  

\subsection{Experimental setting}
We instantiated our technique with EvoSuite, version 1.2.0, and PIT, version 1.8.1. In Section~\ref{sec:technique} we have already described the configuration of EvoSuite with respect to the fitness functions, the time budget and the generation of assertions, and  the mutation operators used with PIT. 

We discriminated inconclusive testability estimates by determining, for each subject method, the threshold for the minimal number of lines of code that we must sample with testability evidences out of the lines for which PIT identified at least a mutant. We computed the thresholds by referring to the classic approximation to the hypergeometric distribution~\cite{krejcie_determining_1970}, setting the confidence level set to 95\%, the population portions to 0.5 and the corresponding accuracy to 15\%.

To compare the performance of our testability estimates with the performance of the  estimates that can be done with traditional software metrics we used the tool CK\footnote{the tool CK is available at \url{https://github.com/mauricioaniche/ck}}~\cite{aniche-ck} to collect the 7 metrics \textit{Loc}, \textit{Rfc}, \textit{Cbo}, \textit{Fan-out}, \textit{Fan-in}, \textit{Cbo-modified} and \textit{Wmc}, for each subject method.
\textit{Loc} is the number of lines of code in the method. \textit{Rfc} is the number of unique method invocations done within the method. \textit{Cbo} is the number of non-primitive data types used in the method.  \textit{Fan-out} is the number of unique classes on which the method depends via method calls.
\textit{Fan-in} is the number of other methods that call the method within the same class. \textit{Cbo-modified} is the sum of \textit{Fan-out} and \textit{Fan-in}.  \textit{Wmc} is the number of branch instructions within the method or 1 for no branch.

\subsection{Results}
\subsubsection{Conclusiveness (RQ1)}\label{sec:RQ1}
Table \ref{tab:nomCovered} reports, for each of the three Java projects (column \textit{Project}) and set of subject methods (column \textit{Subjects}), 
the number of methods for which we achieved conclusive estimations (column \textit{Conclusive}) and the corresponding portion (column \textit{Portion}).  

\begin{table}[htbp]
\caption{\label{tab:nomCovered}Conclusive testability estimations}
\begin{center}
\begin{tabular}{crrr}
\scriptsize
\textbf{Project}& \textbf{Subjects}& \textbf{Conclusive} & \textbf{Portion}\\
\hline
\textit{JFreeChart} & 246 & 206 & 84\%\\
\textit{Closure Compiler} & 111 & 68 & 61\%\\
\textit{OpenMRS} & 241 & 141 & 59\%\\
\end{tabular}
\end{center}
\end{table}

The portion of inconclusive estimations is evidently not negligible, ranging between 16\% and 41\% across the three Java projects. The inspection of the methods with inconclusive estimations revealed that, as we expected, many subject methods were not hit with any test case from EvoSuite since they depended on inputs that EvoSuite cannot generate due to limitations of its current implementation. For example, we identified several methods that take files and streams as inputs (e.g., parameters of type \textit{ObjectInputStream}) that EvoSuite does not currently handle.

We remark that EvoSuite is a research prototype, though very popular in the community of researchers that work on test generation, and we did not expect it to be perfect.
Tuning our technique with further test generators or even ensembles of test generator (as well as experiencing with further mutation analysis tools other than PIT) is an important milestone for our technique to make its way to practice, and definitely the most relevant next goal in our research agenda.
But we also underline the importance of studying the merit of our novel proposal for the cases in which we could indeed achieve conclusive results with the current implementation, which admittedly is our main objective in this paper.

\subsubsection{Correlation with Test Case Complexity (RQ2)}
For the research questions RQ2 and RQ3 we focused on the subject methods for which our technique yielded conclusive results. 

\begin{table*}
\caption{Correlations between testability estimates, static metrics and test case complexity}
\label{tab:comparisonCorr}
\scriptsize\begin{center}
\setlength{\tabcolsep}{4pt}
\begin{small}Legenda: T=Testability, L=Loc, R=Rfc, C=Cbo, CM=CboModified, FI=FanIn, FO=FanOut, W=Wmc, RfcTest=Rfc on test cases.\end{small}\\~\\
\begin{tabular}{r|
cccccccc|
cccccccc|
cccccccc}

&\multicolumn{8}{c|}{\textbf{JFreeChart}}&\multicolumn{8}{c|}{\textbf{Closure Compiler}}&\multicolumn{8}{c}{\textbf{OpenMRS}} \\ 
 & T & L & R & C & CM & FI & FO & W & T & L & R & C & CM & FI & FO & W & T & L & R & C & CM & FI & FO & W \\
\hline 
Loc & -0.16 & & & & & & & & -0.25 & & & & & & & & -0.28 & & & & & & & \\ 
Rfc & -0.44 & 0.64 & & & & & & & - & 0.57 & & & & & & & -0.42 & 0.72 & & & & & & \\ 
Cbo & -0.49 & 0.41 & 0.66 & & & & & & - & 0.33 & 0.69 & & & & & & -0.33 & 0.40 & 0.49 & & & & & \\
CboModified & -0.40 & 0.51 & 0.71 & 0.66 & & & & & -0.31 & 0.24 & 0.43 & 0.36 & & & & & -0.27 & 0.43 & 0.60 & 0.28 & & & & \\ 
FanIn & - & - & -0.20 & -0.28 & 0.15 & & & & - & - & - & - & 0.56 & & & & 0.20 & - & - & -0.30 & 0.38 & & & \\ 
FanOut & -0.43 & 0.50 & 0.76 & 0.77 & 0.93 & -0.16 & & & -0.29 & 0.39 & 0.85 & 0.72 & 0.55 & - & & & -0.44 & 0.51 & 0.78 & 0.57 & 0.74 & -0.19 & & \\ 
Wmc & -0.18 & 0.85 & 0.51 & 0.34 & 0.48 & - & 0.45 & & - & 0.82 & 0.45 & - & - & - & 0.26 & & - & 0.77 & 0.51 & 0.32 & 0.36 & - & 0.43 & \\ 
\hline 
RfcTest & -0.51 & 0.21 & 0.50 & 0.45 & 0.44 & - & 0.49 & 0.16 & -0.41 & - & - & - & 0.23 & - & 0.30 & - & -0.45 & 0.33 & 0.47 & 0.48 & 0.33 & -0.17 & 0.53 & -\\ 
\end{tabular}
\end{center}
\end{table*}

Table \ref{tab:comparisonCorr} reports the correlation (as the Spearman rank correlation coefficient\footnote{The Spearman rank correlation coefficient indicates the extent to which the ranking of the subjects with respect to an indicator produces a good approximation of the ranking  with respect to the other indicator. 
The table also reports the correlation between the testability estimations and the 7 static software metrics that we measured with the tool CK, and the correlation of those 7 metrics between them and with \textit{RfcTest}. The correlation value ranges between -1 and 1, being 1 an indication of perfect correlation (same ranking), -1 and indication of perfect anti-correlation (same inverse ranking) and 0 an indication of no correlation (completely different ranking).})
between our testability estimations, the 7 static software metrics that we measured with the tool CK, and the development complexity of the test cases (measured as the metric \textit{RfcTest}) for the subjects methods in each considered Java project. 
Each cell in the table represents the correlation between the metrics indicated in the titles of the corresponding column and row, respectively. For example, the column \textit{T} represents the correlations between our testability estimates and all other metrics, and the row \textit{RfcTest} represents the correlation of all possible metrics (including our testability estimates) with the development complexity of the cases.
All reported correlation values were computed with R. The missing correlation values (indicated as \emph{dash} symbols in the table) refer to cases for which we did not find support for statistical significance (p-values greater  than 0.05).

We observe that:
\begin{itemize}
\item our testability estimations have a moderate correlation with \textit{RfcTest} for the sets of subject methods of all the considered projects (\textit{JFreeChart}: 0.51,  \textit{Closure Compiler}: 0.41 and \textit{OpenMRS}: 0.45). 
\item Our testability estimates yielded the best correlation with \textit{RfcTest} for the methods of \textit{JFreeChart} and \textit{Closure Compiler}, and the fourth best correlation  for the methods of \textit{OpenMRS}.
\item Our testability estimates have weak correlation with the size of the methods measured as the lines of code (top-left correlation value, row \textit{Loc}, in the three value sets in the table). 
\item The other static metrics resulted in significantly higher correlations with \textit{Loc} (columns \textit{L} in the table) than \textit{Testability}, with the only exceptions of 
\textit{CboModified} in project \textit{Closure Compiler} (where however \textit{CboModified} has only a weak correlation with \textit{RfcTest}).
\end{itemize}

In summary the findings confirm that our testability estimates may contribute to explain the variability in the complexity of the test cases,
while capturing a different phenomenon than the size of the software.
The other software metrics also correlate with the test complexity, sometimes with comparable strength as our testability  estimates, but their independence from  \textit{Loc} is questionable.
Overall, these findings motivate us to explore the possible synergies between our testability estimates and the static metrics.

\subsubsection{Synergy with Static Software Metrics (RQ3)}
We evaluated the performance of the 7 testability indicators obtained by combining each static software metrics with our testability estimates. For each static metric, we obtained the combined  indicator as the average ranking of the two rankings yielded by the static metric and our testability estimates, respectively, for the subject methods.
For the static metrics that are anti-correlated with the testability estimates (all but \textit{FanIn}, see Table \ref{tab:comparisonCorr}) we reversed the testability rankings before computing the combined indicators.

In this study we considered also the methods for which our technique resulted in inconclusive estimates. Since the static metrics are generally available for all methods, and we aim to evaluate if we can benefit from the static metrics in combination with the testability estimates, it makes sense to include those methods as well. For the methods with inconclusive testability estimates, we obtained the combined indicators as just the ranking value yielded by the static metrics (that is, without any additional benefit from testability estimates).

\begin{table}[htbp]
\caption{\label{tab:combinedCorr}Correlation with the combined testability indicators}
\begin{center}
\scriptsize
\begin{tabular}{r|rr|rr|rr}
&\multicolumn{2}{c|}{\textbf{JFreeChart}}&\multicolumn{2}{c|}{\textbf{Closure Compiler}}&\multicolumn{2}{c}{\textbf{OpenMRS}} \\
& base & combined  & base & combined  & base & combined \\
\hline
Loc & 0.17 & 0.34 & - & - & 0.39 & 0.47\\ 
Rfc & 0.46 & 0.53 & 0.23 & 0.32 & 0.52 & 0.55\\ 
Cbo & 0.45 & 0.53 & 0.26 & 0.33 & 0.46 & 0.51\\ 
CboModified & 0.46 & 0.54 & - & 0.35 & 0.39 & 0.52\\ 
FanIn & - & -0.38 & - & -0.31 & -0.20 & -0.32 \\ 
FanOut & 0.50 & 0.58 & 0.23 & 0.30 & 0.51 & 0.53\\ 
Wmc & 0.14 & 0.32 & - & - & 0.28 & 0.43\\ 
\end{tabular}
\end{center}
\end{table}

Table \ref{tab:combinedCorr} reports the correlation between \textit{RfcTest} and the 7 combined testability indicators (columns \textit{combined}) in comparison with the correlation obtained with respect to the base static metrics alone (columns \textit{base})  for the subjects methods in each considered Java project. We report only the correlation values supported with statistical significance (p-value less than 0.05).
The data in the table  show that the correlation yielded with the combined indicators  consistently outperformed the correlation yielded with the corresponding static metrics alone, in most cases with relevant deltas.
This confirms our main research hypothesis that our testability estimates capture a complementary dimension of testability with respect to the traditional software metrics, and can be synergistically combined with those metrics for the purpose of predicting software testability.

\subsection{Threats to validity}
The main threats to the internal validity of our experiments depend on our current choices about the test generation and mutation analysis tools (EvoSuite and PIT) embraced in our current prototype. 
On one hand, our results directly depend on the effectiveness of those tools in sampling the execution space and the fault space of the programs under test, respectively, and thus we might have observed different results if we had experienced with different tools. On the other hand, our experiments suffered of several subjects for which PIT failed to identify  sufficient sets of mutants (the subjects that belonged to the subsets \textit{Tests} in Table~\ref{tab:overallStats}, but that we excluded from the considered subsets \textit{Subjects}) and EvoSuite failed to provide sufficient test cases (the subjects that resulted in inconclusive estimates, see Table~\ref{tab:nomCovered}).

We mitigated the possible threats by focusing our analysis only on the methods that could be reasonably handled with PIT, and by explicitly pinpointing the methods for which EvoSuite allowed us to compute conclusive results. We studied the performance of our technique both as the extent of correlation of our conclusive estimates with the development complexity of the test cases, and by looking into how well our estimates can combine with traditional software metrics also with consideration of our inconclusive results.
But we are aware that we cannot make any strong claim on the efficiency of our current implementation of the technique that we propose in the paper, and  in particular on its specific characteristics of being based on EvoSuite and PIT. Our current claims are only on having provided initial empirical evidence that  
\begin{inparaenum}[(i)]
\item our approach captures a different testability dimension than the size of the software, and
\item it can complement traditional software metrics to reason on software testability in synergistic fashion.
\end{inparaenum}

As for the external validity, our findings may not generalize to other software projects other than the ones that we considered or to programming languages other than Java. In the future, we aim to replicate our experiments on further projects and implement our technique for additional programming languages.

%% file: sections/relatedWorks.tex
The notion of software testability has been first introduced by Freedman~\cite{freedman_testability_1991} along with the related concepts of \emph{observability} and \emph{controllability}.
In turn, these two concepts were inherited from the fields of dynamic systems~\cite{kalman_topics_1969} and hardware testing~\cite{mccluskey_logic_1986}.
Then over time the problem of measuring software testability has been addressed with two classes of approaches, based on either \emph{fault-sensitivity}, which addresses testability by estimating the probability of revealing faults, or \emph{software metrics}, which estimate testability by conjecturing the correlations between software metrics and the testing effort. 

\emph{Fault-sensitivity approaches} were popular in the 90s, with the work about the PIE (or RIP) model~\cite{voas_improving_1991,voas_predicting_1991,voas_pie_1992,voas_software_1995,bologna_object-oriented_1996,demillo1991constraint}. PIE stands for propagate, inject and execute, which are the three main stages of the fault-revealing executions that must be considered to estimate testability. In practice, sensitivity analysis injects simulated faults into the code and evaluates their effect on the outputs.
Bertolino and Strigini exploited this notion of testability to study the relation between testability and reliability~\cite{bertolino_use_1996}.
Lin et al.\ proposed to use a modified version of the PIE technique, which analyzes the structure of the code instead of executing test cases~\cite{lin_estimated_1997}.
Zhao proposed a metric that quantifies the portion of a test suite that can detect specific faults under given test criteria~\cite{zhao_new_2006}.

The strength of the fault-sensitivity approaches is to refer to actual faults, the weakness is on the performance side, since the number of input data that need to be provided for sensitivity analysis is high even for small programs. For these reasons, researchers progressively moved to software metrics, which are considered more cost-effective to compute. 
Our work however shows that dynamic measures derived from observability and controllability  are important factors not subsumed by static software metrics, which should be rather considered in combination with them.  

\emph{Software metrics} derive testability indexes from metrics that capture information about the static structure of the code. These studies aim at finding a correlation between the static metrics and the testability of the analyzed software, to identify which metrics are best predictors of testability. Different research efforts studied different combination of metrics. 
Khalid et al.\ proposed static metrics that aim at  estimating the complexity of an object and evaluate their performance for testability prediction~\cite{khalid_analysis_2010}. Alshawan et al.\ proposed a set of static metrics specific to web applications~\cite{alshahwan_improving_2009}. A large body of papers refer to the so called CK metrics for object oriented software~\cite{chidamber_metrics_1994}.
Gupta et al.\ propose a fuzzy approach to integrate the CK metrics in a unique metric that should represent the testability~\cite{gupta_fuzzy_2005}.  Singh et al.\ and Zhou et al.\ used neural networks and linear regression, respectively, to predict testability to combine several software static metrics to predict testability~\cite{zhou_-depth_2012,singh_predicting_2008}. 

All research efforts share the challenge of deriving a ground truth for evaluating the goodness of the proposed techniques. 
Typically researchers referred their experiments to metrics that quantify the testing effort as the size or the complexity of test suites available in software repositories. Possible metrics include: the number of test cases, the number of lines of test code, the number of assertions, the number of all or unique method calls in test cases, and the average cyclomatic complexity of the test cases. For instance, Bruntik and van Deursen studied the correlation between object-oriented metrics and the testing effort estimated as above~\cite{bruntink_predicting_2004,bruntink_empirical_2006}. Other studies measured the test effort as the time required for completing the testing tasks~\cite{gupta_fuzzy_2005}. This datum is however seldom available as historical data in software repositories. Others referred to code coverage to evaluate testability indicators with respect the quality of the test suite.
Terragni et al.\ referred to coverage data normalized with respect to the size of the test cases~\cite{terragni_measuring_2020}. In line with previous studies, we consider the complexity of the test cases (measured as RfcTest) as ground truth of testability.

%% file: sections/conclusions.tex
In this paper we discuss a new approach for measuring software testability. Our approach tackles the testability measurement problem explicitly, by operationally estimating the degree of controllability and observability of a software component. In particular, our approach samples the test space and the fault space of the target component, therein collecting empirical evidence of the easiness or hardness to accomplish effective testing. Our approach provides a novel and direct way of dealing with testability, compared to previous work that attempts to measure testability based on arguable relations with code size, code complexity and fault sensitivity. We computed our metric for Java methods and performed experiments with 598 subject methods from 3 Java projects. The results show that our approach captures a testability dimension that static metrics do not well represent, and thus it can well complement traditional software metrics. 

In future work we aim to address the drawbacks that we discussed in the paper about the automatic testing and mutation analysis tools, aiming to refine and  expand the empirical evidence on the effectiveness of the proposed approach. In addition, we believe that the unexecuted and unrevealed mutants that our technique pinpoints may provide concrete examples of untestable software behaviors that can be of interest for the engineers to understand the reasons of low testability, and we thus aim to investigate this different type of exploitation of our approach.